\documentclass[aps,pra,twocolumn,groupedaddress,epsfig,amsmath,floatfix]{revtex4}
\usepackage{epsfig,amsmath}
\usepackage{graphicx}
\usepackage{dcolumn}
\usepackage{bm}

\begin{document}

\def\ao#1{{ Appl.\ Optics} {\bf #1}}
\def\opex#1{{ Optics\ Expr.} {\bf #1}}
\def\josab#1{{ J. Opt.\ Soc.\ Am.\ B\/} {\bf#1}}
\def\oe#1{{ Opt.\ Express} {\bf#1}}
\def\oc#1{{ Opt.\ Commun.} {\bf#1}}
\def\jpb#1{{ J.\ Phys.\ B} {\bf#1}}
\def\ajp#1{{ Am.\ J.\ Phys.} {\bf#1}}
\def\pra#1{{ Phys.\ Rev. A\/} {\bf#1}}
\def\pre#1{{ Phys.\ Rev. E\/} {\bf#1}}
\def\prl#1{{ Phys.\ Rev.\ Lett.} {\bf#1}}
\def\ave#1{{\langle}{#1}{\rangle}}

\title{Non-Sequential Double Ionization is a Completely Classical 
Photoelectric Effect}

\author{Phay J. Ho,$^1$ R. Panfili,$^{1,2}$ S.L. Haan,$^3$ and J.H. Eberly$^1$}

\affiliation{{$^1$Department of Physics and Astronomy,}
{University of Rochester, Rochester, NY 14627,}\\
{$^2$Spectral Sciences, Inc., Burlington MA 01803,}
{$^3$Department of Physics and Astronomy, Calvin College, Grand Rapids MI 49546}}

\begin{abstract}
We introduce a unified and simplified theory of atomic double 
ionization.  Our results show that at high laser intensities ($I \ge 
10^{14}$ watts/cm$^2$) purely classical correlation is strong enough 
to account for all of the main features observed in experiments to 
date.
\end{abstract}

\maketitle

Short-pulse lasers with high peak intensities ($10^{14} \le I \le 10^{16}$, in watts/cm$^2$) now produce multiphoton generation of double ionization, the two-electron photoelectric effect, with surprising results. To summarize briefly, the experimental data show that two atomic (or molecular) outer-shell electrons are highly correlated when photo-ejected, with a double ionization rate that can be 1-million times higher than uncorrelated sequential theory \cite{PPT-ADK} allows, so the process is called non-sequential double ionization (NSDI).  The first laboratory results were reported in 1992 and 1993 \cite{fittinghoff-etal, walker-etal}, showing an anomalously high double ionization yield, the principal experimental signature of NSDI. Additional data is being reported from momentum spectroscopy experiments \cite{weber_corr, weber_He, moshammer_Ne, weber_Ar, feuerstein_Ar, eremina, DiMauro01, weckenbrock_Ne}. The momentum distribution data, along with the ion-yield data, serve as the benchmarks for various theoretical models.

The mechanism that makes NSDI correlation so effective is far from settled, and theoretical exploration has been extensive \cite{Becker-Faisal1, coulomb-focusing, Goreslavski, softcollision, Panfili-etal02, Texas}.  Almost all existing calculations refer to, or are closely guided by, a single few-step rescattering model \cite{two-step, corkum}, which is based on an imagined picture in which one electron escapes the atom by quantum tunneling through a field-lowered barrier and is then phase-coherently and classically forced by the laser away from and then back to the core where a quantum collision liberates both electrons at once (consistent with the term non-sequential). However, the patchwork of {\em ad hoc} elements typically employed \cite{Taylor} has not been claimed to make a complete, i.e., self-contained, theory.  It is the purpose of this note to show that a self-contained theory exists that is compatible with essentially all prominent features of NSDI.

Our theory is dynamically classical, and discards all aspects of quantum mechanics including tunneling. It is built on the need for strong electron correlation to explain NSDI, and so must be intrinsically a two-electron theory. We do not advocate such a theory for an electron that does not have the advantage of a strongly correlated partner. It turns out that entirely classical interactions are  adequate to generate very strong two-electron correlation, as observed in NSDI, and quantum theory is not needed. Of course atoms are quantum objects but in such strong fields as are used for NSDI it is mainly electron physics rather than atomic physics that determines the experimental outcome. In this sense the early remark of Corkum \cite{corkum} advocating the adoption of a plasma perspective was quite appropriate.

\begin{figure}
\centerline{\includegraphics[width=2.8in]{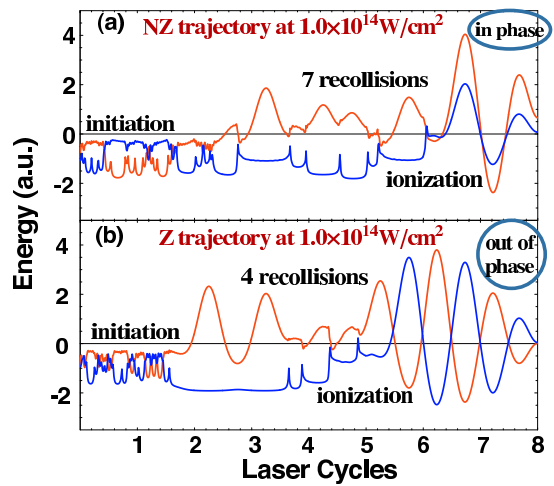}}
\caption{Energy vs. time plots showing 4 distinct stages of NSDI. 
The orange and blue lines track the energies of two electrons. After 
repeated exchanges of energy while both are bound, one electron 
(orange) acquires positive energy but returns a number of times to 
the core. Its strong recollisions show up as spikes in the energy 
line of the still-bound inner electron. After several spikes the 
phasing is right to liberate the inner electron and the doubly 
ionized pair exhibit free-electron jitter motion.}
\label{fig.4-stage}
\end{figure}

We note that a new form of energy analysis is very helpful. The graph in Fig. \ref{fig.4-stage} shows the sum of kinetic energy, electron-nucleus binding energy, e-e correlation energy and laser field interaction energy for each of two electrons during the laser pulse.  Such Newtonian energy trajectories universally display the same sequence of stages, which we propose to accept as the dynamical signature of a high-field double-ionization event.

\begin{figure}
\centerline{\includegraphics[width=2.4in]{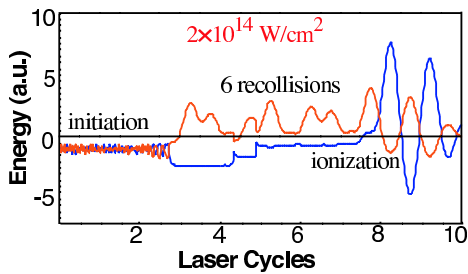}}
\caption{Classical high-field energy trajectories calculated in three 
dimensions, showing the same characteristic four stages of NSDI, as in Fig. \ref{fig.4-stage}. }
\label{fig.energies3d}
\end{figure}

For simplicity we have made most of our classical calculations with a one-dimensional model, but the results don't depend strongly on this. The results shown in Fig. \ref{fig.4-stage} were calculated one-dimensionally but a fully three-dimensional calculation displays exactly the same sequence of stages, as shown in Fig. \ref{fig.energies3d}. A quick inspection of these energy plots shows that we can refer to the stages of high field double ionization as initiation, recollision, ionization and jitter. This characterization is not far from that conjectured by Yudin and Ivanov \cite{softcollision} on the basis of selected Newtonian trajectories for a single electron. Our pictures show details in a time domain not seen before (2e dynamics prior to single ionization), and they also reveal a previously unremarked electron-pair phasing. Below we correlate this phasing with momentum properties of the electron-ion products in an intuitively appealing way.

One can easily describe the characteristic features of each of the four NSDI stages as follows. \\ 
$\bullet$ Initiation stage: Both electrons are confined in bound orbits in the nuclear potential and many rapid e-e interactions occur. One electron can easily take energy from the other, and escape without need for tunneling. \\ 
$\bullet$ Recollision stage: The semi-liberated electron returns to the core repeatedly. During each return the efficiency of energy transfer rests on the relative motional phase between the two electrons. \\ 
$\bullet$ Ionization stage: A final collision leads to highly correlated double ionization, which we find usually occurs after several recollisions, not the first.\\
$\bullet$ Jitter stage: The two electrons exhibit the jitter oscillations characteristic of free electrons. The oscillations are exactly in phase or out of phase with each other.

\begin{figure}
\centerline{\includegraphics[width=1.3in]{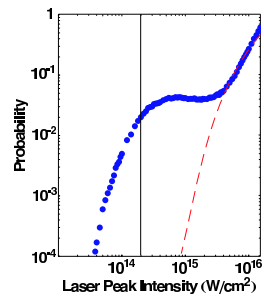}}
\caption{The ``knee" signature of NSDI data is clearly predicted even 
in a classical calculation. The dashed curve (red) is inserted by 
hand to indicate the prediction made by sequential quantum tunneling 
theory \cite{PPT-ADK}.}
\label{fig.NSDI-ion-curve}
\end{figure}

We note that in an entirely classical picture one is able to work with a self-contained and almost completely unified theory. That is, in contrast to mixed classical-quantum patchworks, there are few adjustable parameters, and {\em ad hoc} decisions about timings, cross sections, matrix elements, etc., are not needed. Moreover the calculations are complete in the sense that they begin when the field turns on and continue without modification to whatever later time is of interest, and are exact in the sense that perturbation theory plays no role. A classical prediction of double ionization might at first be considered accidental, except that Newtonian calculations produce the anomalous high-yield ``knee", the principal signature of NSDI, as shown in Fig. \ref{fig.NSDI-ion-curve}, where the vertical line shows the location of the intensity threshold predicted by the old two-step theory. Moreover, we find no strongly correlated double-electron effect that needs quantum mechanics in order to be understood, not even the initial liberation of the first electron, which in the two-step model is assumed to originate in quantum tunneling (remarkably, the need for this Ansatz seems never to have been tested for two-electron phenomena).

Our classical method is straightforward once the ensemble of initial conditions is described (for this, see \cite{Panfili-etal01}). We employ a large microcanonical ensemble (100,000 - 500,000 members) of independent two-electron atoms, each with initial total energy equal to the energy of the corresponding two-electron quantum ground state. Since NSDI is understood as a universal high-field phenomenon to be found in essentially any multi-electron atomic species we try to make non-specific calculations as far as possible. One  nonspecific component is the familiar quasi-Coulomb model for the one-dimensional electron-electron and electron-nucleus potentials \cite{Javanainen-etal87} $V(x) = -1/\sqrt{x^2 + a^2}$, or its analog in three-dimensional calculations. When we take the  soft-core parameter the same, $a=1$, for all the interactions, the initial two-electron energy is fixed at the ground-state energy -2.24 a.u.  We choose an 8-cycle or 10-cycle (25-30 fs) sinusoidal laser pulse with the wavelength 780 nm (frequency $\omega$ = 0.0584 a.u.) and a trapezoidal envelope. By integrating the Newtonian equations of motion we can then numerically follow any two-electron trajectory and the ensemble of them is suited to a statistical analysis that can be compared with experimental results and with the results of other theoretical models.

\begin{figure}
\centerline{\includegraphics[width=2.8in]{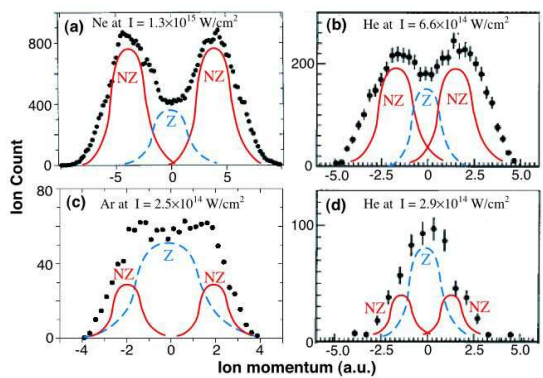}}
\caption{Final ion momentum distributions of neon \cite{moshammer_Ne}, helium \cite{weber_He} and argon \cite{weber_Ar} measured using COLTRIMS.  Panels (b) and (d) show the momentum distribution of He ions at different laser intensities.  Two solid curves (red) and a dashed curve (blue) are inserted by hand under the experimental data to identify contributions from the two types of NSDI trajectory.  One type gives distributions peaked at Zero (Z) momentum, and the other gives distributions with symmetrically placed Non-Zero (NZ) momentum peaks.  The relative sizes of the Z and NZ peaks in each distribution depend on the atomic species and the laser intensities.}
\label{fig.COLTRIMS}
\end{figure}

The final momentum distribution of NSDI ions of Ne, He and Ar atoms measured using COLTRIMS are different, as shown in panels (a), (b) and (c) in Fig. \ref{fig.COLTRIMS}.  The distribution of Ne shows a double-peak structure with a valley at zero \cite{moshammer_Ne,eremina}, the distribution of He shows a double-peak structure with its valley nearly filled \cite{weber_He}, and the distribution of Ar shows a broad peak centered at zero \cite{weber_Ar, feuerstein_Ar, eremina}.  One can interpret that there are two groups of ions or trajectories: one group gives a distribution of ions peaked at Zero (Z) momentum, and the other gives two distributions of ions peaked at Non-Zero (NZ) momenta.  These two base distributions superpose to give the observed distributions.  In Ne, the main contribution comes from the NZ ions with minimal contribution from the Z group ions.  In He, the main contribution also comes from the NZ ions, but it has more contribution from Z ions than in Ne in order to fill the valley between the double peaks.  As for Ar, the number of NZ ions is significantly less than the Z ions to give the broad peak structure centered at zero.  This interpretation is consistent with the analysis by Feuerstein et al. \cite{feuerstein_Ar}, who separate the distributions of NZ and Z ions in $Ar^{2+}$ with Z ions as the majority.  These two groups of ions were previously identified in our classical simulation \cite{Ho-Eberly03}.

A new result obtained from our purely classical treatment of strongly correlated electron dynamics, as revealed by the new energy trajectories, is the first intuitively natural explanation for a main component of recoil ion momentum distributions such as are obtained experimentally for helium, neon and argon.  Figs. \ref{fig.4-stage} and \ref{fig.energies3d} show that the end stage of NSDI finds the two ejected electrons exhibiting jitter oscillations that are either exactly in phase or out of phase.  These energy oscillations are due primarily to the potential energy from interaction with the laser field.  The in-phase oscillations have the electrons typically escaping the binding potential in the same half laser cycle after the final electron-electron collision, although some of them can have a lag time of an even number of laser cycles.  This situation  gives relatively non-zero sum of electron momentum and is denoted above as the non-zero (NZ) recoil case.  On the other hand, the out of phase events find the second electron to be field-ionized (rather than directly collision-ionized) in an odd half-cycle after the first one departs.  The electrons emerge on opposite sides of the nucleus.  This situation gives zero or relatively small sum of electron momentum and is denoted as the zero (Z) recoil case. 

\begin{figure}
\centerline{\includegraphics[width=3.2in]{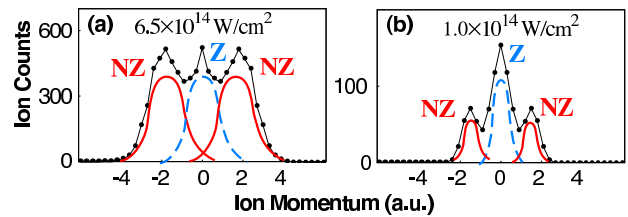}}
\caption{Final ion momentum distributions calculated classically 
using the classical ensemble method.  The letters Z and NZ label the peaks of 
the broad but distinct groups of trajectories with small or Zero ion 
momentum and substantially Non-Zero ion momentum respectively.  }
\label{fig.NSDI-pdist}
\end{figure}

For completeness, Fig. \ref{fig.NSDI-pdist} shows the momentum distributions arising from classical calculations, with peaks labelled either Z or NZ.  Classical dynamics predicts that as the laser intensity decreases the NZ fraction and the width of the momentum distribution decrease.  These two effects have been observed experimentally \cite{weber_He, weber_Ar,eremina} (compare Figs. \ref{fig.COLTRIMS} (b) and (d) to see these effects).  In addition, our classical calculation finds electrons with energy above $2 U_p$, where $U_p=I/4\omega^2$ is the ponderomotive energy .  The number of these electrons is significantly less than the number of electrons with energy below $2 U_p$.  These two findings are consistent with the experimental observations on helium, neon and argon from DiMauro's group \cite{walker-etal,DiMauro01}.  The details of the analysis will be presented elsewhere. 

Before concluding, it is appropriate to explain what is gained in a classical picture. The biggest advantage is calculational, since Newtonian dynamics can be followed in full detail, something out of the question for an approach via time-dependent two-electron Schr\"odinger theory.  On the conceptual side the advantages are also substantial and significant. Disconnected {\em ad hoc} elements disappear, and the theory is unified. We note that in the two-step picture and its extensions the need for quantum tunneling is an {\em ad hoc} assumption that has not been critically examined. The justification to patch quantum tunneling onto classically forced motion requires another {\em ad hoc} Ansatz. A third {\em ad hoc} element occurs in analyses that have a classically returning ``outer" electron obey a quantum collision cross section when it encounters the nucleus and an artificially quiescient ``inner" electron. A more sophisticated example occurs in calculations using a selection of S-matrix elements guided only by the two-step picture. The negative effect here is that the rationale of S-matrix theory is abandoned, i.e., the next higher order of approximation cannot be identified, let alone calculated.  The {\em ad hoc} elements just mentioned are almost all inherited from a one-active-electron approach that has worked extremely well in single ionization contexts. The work of Lewenstein, et al., \cite{Lewenstein-etal} clearly demonstrates this. However a one-electron theory is not suitable for an intrinsically two-electron phenomenon.

In summary, we have made an entirely classical study of the response of two-electron atoms to intense short-pulse laser radiation in the very high-field regime. Our purpose is not to find agreement with NSDI data specific to any real atom, but to determine whether it is reasonable to assign the strong electron correlation associated with NSDI to classical orgins. This is clearly the case and the consequences are important because a straightforward and numerically easy route is now opened up for undertaking systematic and wide-ranging exploration of time-dependent and phase-coherent multi-electron effects in a strong radiation field. This is more than academically interesting because pictures of electrons in such trajectories are being used to guide important initiatives in atomic, molecular and optical physics, including attosecond timing of atomic and molecular processes \cite{ATTO}, the generation of controlled intramolecular single-electron beam currents \cite{E-BEAM} and short-wavelength coherent radiation \cite{X-RAY}, and the use of strong short-pulse laser fields to control electron motion in general \cite{CONTROL}. In particular, we have shown how central is the role of e-e dynamical interaction in the ionizing dynamics.  The strength of the e-e interaction has previously been undervalued, but it dynamically facilitates the required energy exchange between electrons.  Different multiple-recollision channels produce trajectories with different final momenta, producing similar or opposite jitter phasings, which underlie the experimental NZ and Z momentum distributions, respectively.  Most importantly, we have shown that the major signatures of NSDI, at least those that are observed experimentally, can be understood using classical physics, leaving only a minor (or possibly future) role for quantum effects to play. It is intriguing to guess that experimenters have already been registering some three-electron effects with NSDI observations without knowing it.

Acknowledgements: This work was supported by NSF grants PHY-0072359 to University of Rochester and PHY-0355035 to Calvin College.  We want to acknowlegde contributions of L. Breen and D. Tannor for Fig. \ref{fig.energies3d}.  Also, we have benefitted over an extended period from discussions on this topic with W. Becker, P.B. Corkum, L.M. DiMauro, M. Yu. Ivanov, J. Marangos, W. Sandner and D. Zeidler.

\bibliography{apssamp}

\begin{thebibliography}{99}


\bibitem{PPT-ADK} See A. M. Perelomov, V. S. Popov, and M . V. Terentev, Sov. Phys. JETP {\bf 23}, 924 (1966), and M. V. Ammosov, N. B. Delone and V. P. Kra\u{i}nov, {\it{Sov. Phys. JETP}} {\bf 64}, 1191 (1986).

\bibitem{fittinghoff-etal} D. N. Fittinghoff, P. R. Bolton, B. Chang and K. C. Kulander, {\it{Phys. Rev. Lett.}} {\bf 69}, 2642 (1992).

\bibitem{walker-etal} B. Walker {\it{et al.}}, {\it{Phys. Rev. Lett.}} {\bf 73}, 1227 (1994).

\bibitem{weber_corr} Th. Weber {\it{et al.}}, Nature (London) {\bf 405}, 658 (2000).

\bibitem{weber_He} Th. Weber {\it{et al.}}, {\it{Phys. Rev. Lett.}} {\bf 84}, 443 (2000).

\bibitem{moshammer_Ne} R. Moshammer {\it{et al.}}, Phys. Rev. Lett. {\bf 84}, 447(2000).

\bibitem{weber_Ar} Th. Weber {\it{et al.}}, {\it J. Phys. B} {\bf 33}, L127 (2000).

\bibitem{feuerstein_Ar} B. Feuerstein {\it{et al.}}, {\it{Phys. Rev. Lett.}} {\bf 87}, 043003 (2001).

\bibitem{eremina} E. Eremina {\it{et al.}}, {\it J. Phys. B} {\bf 36}, 3269 (2003).

\bibitem{DiMauro01} B. Sheehy, {\it{et al.}}, {\it{Phys. Rev. A}} {\bf 58}, 3942 (1998); R. Lafon {\it{et al.}}, {\it{Phys. Rev. Lett.}} {\bf 86}, 2762 (2001).

\bibitem{weckenbrock_Ne} M. Weckenbrock {\it{et al.}}, {\it{Phys. Rev. Lett.}} {\bf 92}, 213002 (2004).

\bibitem{Becker-Faisal1} A. Becker and F. H. M. Faisal, {\it{Phys. Rev. A}} {\bf 59}, R1742 (1999).

\bibitem{coulomb-focusing} T. Brabec, M. Yu. Ivanov and P. B. Corkum, {\it{Phys. Rev. A}} {\bf 54}, R2551 (1996).

\bibitem{Goreslavski} S.P. Goreslavskii, S.V. Popruzhenko, R. Kopold, and W. Becker, {\it Phys. Rev. A}  {\bf 64}, 053402 (2001).

\bibitem{softcollision} G.L. Yudin and M. Yu. Ivanov, {\it{Phys. Rev. A}} {63}, 033404 (2001).

\bibitem{Panfili-etal02} R. Panfili, S. L. Haan, and J. H. Eberly, {\it{Phys. Rev. Lett.}} {\bf 89}, 113001 (2002), and S. L. Haan, P. S. Wheeler, R. Panfili and J. H. Eberly, {\it{Phys. Rev. A}} {\bf 66}, 061402(R) (2002).

\bibitem{Texas} L.-B. Fu, J. Liu and S.-G. Chen, {\it{Phys. Rev. A}} {\bf 65}, 021406(R) (2002).

\bibitem{two-step} K.C. Kulander, K. J. Schafer and J.L. Krause, in {\em Super Intense Laser-Atom Physics}, B. Piraux, A. L'Huillier and K. Rzazewski, Eds. (Plenum, New York, 1995), p. 95.

\bibitem{corkum} P. B. Corkum, {\it{Phys. Rev. Lett.}} {\bf 71}, 1994 (1993).

\bibitem{Taylor} The Taylor group is an exception to comments about {\em ad hoc} or two-step approaches. They have systematically pursued exact numerical solutions to the time-dependent Schr\"odinger equation for helium in strong laser fields. See J. S. Parker {\it{et al.}}, {\it J. Phys. B} {\bf 36}, L393 (2003) and references therein. However, to the present time the numerical challenges are so great that the helium results cannot be extended to other atoms being studied experimentally.

\bibitem{Panfili-etal01} R. Panfili, J. H. Eberly and S. L. Haan, {\it Opt. Express} {\bf 8}, 431 (2001).

\bibitem{Javanainen-etal87} See J. Javanainen, J.H. Eberly and Q. Su, \pra {38}, 3430 (1987), and Q. Su and J.H. Eberly, \pra {44}, 5997 (1991).

\bibitem{Ho-Eberly03} P.J. Ho and J.H. Eberly, {\it Opt. Express} {\bf 11}, 2826 (2003).

\bibitem{Lewenstein-etal} M. Lewenstein, P. Balcou, M. Y. Ivanov, A. LHuillier ,and P. B. Corkum,  {\it{Phys. Rev. A}} {\bf 49}, 2117 (1994).

\bibitem{ATTO} For example, ``Strong field path control using attosecond pulse trains," K.J. Schafer, M. B. Gaarde, A. Heinrich, J. Biegert and U. Keller, {\it{Phys. Rev. Lett.}} {92}, 023003 (2004).

\bibitem{E-BEAM}  For example, ``Sub-laser-cycle electron pulses for probing molecular dynamics", H. Niikura, {\it et al.}, {\it{Nature}} {\bf 417}, 917 (2002).

\bibitem{X-RAY} For example, ``Coherent Soft X-ray Generation in the Water Window with Quasi-Phase Matching", E. A. Gibson, {\it{et al.}}, {\it{Science}}  {\bf 302}, 95 (2003).

\bibitem{CONTROL} For example, ``Attosecond control of electronic processes by intense light fields", A. Balt\u{u}ska, {\it{et al.}}, {\it{Nature}} {\bf 421}, 611 (2003). \end{thebibliography}

\end{document}